\documentclass[nofootinbib]{revtex4}

\usepackage{listings,comment,xcolor,amsmath,courier}

\definecolor{lgray}{rgb}{.98,.98,.98}

\lstdefinelanguage{maxima}{morekeywords={canform,for,thru,from,do,tex,load,linenum,imetric,igeowedge_flag,components,extdiff,ishow,remcomps,decsym,anti,all,matchdeclare,apply,defrule,diff,contract,expand,apply1,lhs,rhs,idiff,atom,true,map,lambda,covdiff,kdelta},sensitive=t}
\begin{document}

\title{Field theory with the Maxima computer algebra system}

\author{Viktor T. Toth$^1$}

\affiliation{\vskip 3pt $^1$Ottawa, Ontario K1N 9H5, Canada}

\date{\today}

\begin{abstract}
The Maxima computer algebra system, the open-source successor to MACSYMA, the first general-purpose computer algebra system that was initially developed at the Massachusetts Institute of Technology in the late 1960s and later distributed by the United States Department of Energy, has some remarkable capabilities, some of which are implemented in the form of add-on, ``share'' packages that are distributed along with the core Maxima system. One such share package is itensor, for indicial tensor manipulation. One of the more remarkable features of itensor is functional differentiation. Through this, it is possible to use itensor to develop a Lagrangian field theory and derive the corresponding field equations. In the present note, we demonstrate this capability by deriving Maxwell's equations from the Maxwell Lagrangian, and exploring the properties of the system, including current conservation.
\end{abstract}

\maketitle

\section{Introduction}

Both classical and quantum field theories are usually formulated using a Lagrangian density \cite{LL1975}. Examples include the Einstein--Hilbert Lagrangian of general relativity, or the ``monster'' Lagrangian of the Standard Model of particle physics, which includes kinetic and potential terms associated with the entire particle content of the model and all their interactions.

Especially in the classical physics context, the immediate use of a Lagrangian density is by way of deriving the corresponding Euler--Lagrange field equations \cite{LR1989}. This is how, for instance, Einstein's field equations are obtained from the aforementioned Einstein--Hilbert Lagrangian, or Maxwell's equations may be obtained from the Lagrangian of classical electrodynamics.

The process of obtaining the Euler--Lagrange equations of motion is straightforward but often cumbersome and consequently, error-prone. For this reason, computer algebra can bring significant benefits, streamlining the derivation and cutting down on the number of avoidable errors.

Unfortunately, key to the derivation of the Euler--Lagrange equations, functional differentiation, is not always easily achievable within the syntax of existing computer algebra packages.

The Maxima computer algebra system\footnote{See \url{https://maxima.sourceforge.io/} for information on downloading and installing Maxima or accessing its source code.} may serve as a useful exception. Derived from MACSYMA \cite{FATEMAN1979,MOSES2012}, arguably the first general-purpose computer algebra system that was initially developed at the Massachusetts Institute of Technology in the late 1960s and later distributed by the United States Department of Energy, Maxima is a comprehensive, free, and open-source computer algebra system on par with well-known commercial alternatives such as Maple or Mathematica.

In addition to its core capabilities, Maxima has many loadable extensions. One such extension is {\tt itensor}, Maxima's indicial tensor manipulation package\footnote{My participation in Maxima's development began in the early 2000's when I ``inherited'' its tensor packages, which were in a dysfunctional state at the time. The core concepts of {\tt itensor} and {\tt ctensor} are due to their original creators. My contributions included, e.g., functional differentiation. -- VTT} \cite{Bogen1977}. This package has the perhaps unique feature of treating tensors as ``opaque'' indexed objects. The package applies formal rules of algebra and calculus to these objects even as it remains agnostic to the values that these indexed objects represent. This makes {\tt itensor} uniquely suited to deal with field theory problems, including generally covariant formalisms \cite{Toth2005,2018LRR}.

In this paper, we demonstrate the use of {\tt itensor} in field theory applications by explicitly deriving Maxwell's equations from the Lagrangian of the electrodynamic field, which we introduce in Section~\ref{sec:fld}. Next, in Section~\ref{sec:itensor}, we describe the {\tt itensor} package, in particular its features that are relevant to the derivation. We perform the derivation in Section~\ref{sec:derive} and present a complete Maxima program. Our conclusions are presented in Section~\ref{sec:end}.

\section{The Lagrangian density of the electromagnetic field}
\label{sec:fld}

As it is well known, Maxwell's theory can be derived from a postulated Lagrangian density. The main feature of this Lagrangian density is its gauge invariance that, in turn, yields a conserved current.
\begin{align}
{\cal L}_{\rm EM} = -\frac{1}{4}F_{\kappa\lambda}F^{\kappa\lambda}+j_\kappa A^\kappa,
\end{align}
where $F_{\kappa\lambda}=\partial_\kappa A_\lambda-\partial_\lambda A_\kappa$ is the electromagnetic field tensor corresponding to the 4-potential $A_\mu$. (Indices run from 1 through the number of dimensions $D$, and we use the Einstein summation convention, $x_\mu y^\mu=\sum_{\mu=1}^D x_\mu y^\mu$.)

The theory is obviously invariant under the gauge transformation $A_\mu\to A_\mu + \partial_\mu \theta$ where $\theta$ is an arbitrary (but differentiable) ``gauge'' field.

The Euler-Lagrange equation that corresponds to this Lagrangian is written as
\begin{align}
\frac{\partial(
{\cal L}_{\rm EM})}{\partial A_\mu}-\nabla_\nu\frac{\partial(
{\cal L}_{\rm EM})}{\partial(\nabla_\nu A_\mu)}=0,
\end{align}
where $\nabla_\mu$ is the covariant derivative with respect to the coordinate $x^\mu$.

The first term in this equation is trivial: $\partial{\cal L}_{\rm EM}/\partial A_\mu=j^\mu$, as follows directly from the definition of ${\cal L}_{\rm EM}$. As to the second term, we first note that
\begin{align}
F_{\kappa\lambda}F^{\kappa\lambda}=F_{\kappa\lambda}g^{\kappa\alpha}g^{\lambda\beta}F_{\alpha\beta},
\end{align}
and that
\begin{align}
\frac{\partial F_{\kappa\lambda}}{\partial(\nabla_\nu A_\mu)}=\delta^\nu_\kappa\delta^\mu_\lambda-\delta^\nu_\lambda\delta^\mu_\kappa,
\end{align}
thus
\begin{align}
\frac{\partial}{\partial(\nabla_\nu A_\mu)}F_{\kappa\lambda}F^{\kappa\lambda}=(\delta^\nu_\kappa\delta^\mu_\lambda-\delta^\nu_\lambda\delta^\mu_\kappa)F^{\kappa\lambda}+
F^{\alpha\beta}(\delta^\nu_\alpha\delta^\mu_\beta-\delta^\nu_\beta\delta^\mu_\alpha)=4F^{\nu\mu}=-4F^{\mu\nu}.
\end{align}
Therefore
\begin{align}
\nabla_\nu\frac{\partial{\cal L}_{\rm EM}}{\partial(\nabla_\nu A_\mu)}=-\frac{1}{4}\nabla_\nu\frac{\partial}{\partial(\nabla_\nu A_\mu)}F_{\kappa\lambda}F^{\kappa\lambda}=\nabla_\nu F^{\mu\nu}=F^{\mu\nu}_{;\nu},
\end{align}
where we used the common shorthand notation $\nabla_\mu f=f_{;\mu}$ for covariant derivatives. Finally, we can now put everything together and obtain the explicit form of the Euler-Lagrange equation, which is just two of Maxwell's equations, Gauss's law and Amp\`ere's law, combined in generally covariant form:
\begin{align}
j^\mu+F^{\mu\nu}_{;\nu}=0.
\label{eq:Maxwell2}
\end{align}
The remaining two Maxwell equations, Gauss's law for magnetism and Faraday's law, are trivial identities that follow from the nilpotence of the exterior derivative. Given $F_{\mu\nu}=\partial_{[\mu}A_{\nu]}$ where we used square brackets for antisymmetrization, i.e., $X_{[\alpha\beta]}\equiv X_{\alpha\beta}-X_{\beta\alpha}$, we have
\begin{align}
\partial_{[\kappa}F_{\mu]\nu}=\partial_\kappa F_{\mu\nu}-\partial_\kappa F_{\nu\mu}=
\partial_\kappa\partial_\mu A_\nu-\partial_\mu\partial_\kappa A_\nu
-\partial_\kappa\partial_\nu A_\mu+\partial_\nu\partial_\kappa A_\mu=0.
\label{eq:Maxwell1}
\end{align}

All this, of course, is well known, standard theory. How would we go about deriving these equations using computer algebra?

\section{Brief tour of the {\tt itensor} package}
\label{sec:itensor}

The {\tt itensor} package treats tensors as opaque objects, manipulated via their indices. As such, {\tt itensor} is especially suited for computations where general covariance is maintained, such as general relativity. The package has built-in support for the metric, for the Kronecker delta, and for utilizing the symmetry properties of tensors for algebraic simplification. All this is accomplished without the need to define tensor components.

The {\tt itensor} package also has facilities for both indexed tensor calculus, including functional differentiation with respect to tensor quantities. These facilities make it possible to use {\tt itensor} to investigate field theories that are formulated using a Lagrangian density functional.

Indexed objects are represented in {\tt itensor} as ersatz functions. For instance, consider the object {\tt T([a,b],[c,i],i2,i1)}. This expression represents a tensor with two covariant indices ($a, b$), two contravariant indices ($c, i$) and two coordinate derivative indices ($i_1, i_2$): $T_{ab,i_2i_1}^{ci}$.

The {\tt itensor} package has a feature, in the form of the {\tt ishow()} function, to ``pretty print'' indexed tensor expressions. This feature is best used in conjunction with using the dollar sign, {\tt \$}, as the line termination character, to suppress normal output. E.g.\footnote{Note that in this paper, we present Maxima examples as they appear through Maxima's command-line, plain-text interface in a text terminal. Front-ends to Maxima, such as {\tt wxMaxima}, exist that offer rendered output in graphical environments such as Microsoft Windows.},
\begin{lstlisting}[frame=single,xleftmargin=0.25in,xrightmargin=0.25in]
(%i1) load(itensor)$
(%i2) ishow(T([a,b],[c,i],i2,i1))$
                                   c i
(%t2)                             T
                                   a b,i2 i1
\end{lstlisting}

The {\tt itensor} package can do index contraction and understands the metric. In particular, it understands that inner products with the metric tensor raise or lower indices. The package also implements the Kronecker delta. These functionalities can be demonstrated, e.g., by the following:

\begin{lstlisting}[frame=single,xleftmargin=0.25in,xrightmargin=0.25in]
(%i3) imetric(g)$
(%i4) ishow((g([],[a,b])*kdelta([i],[j])*A([b,c,j,k],[l])))$
                              a b  l             j
(%t4)                        g    A        kdelta
                                   b c j k       i
(%i5) ishow(contract(%))$
                                     a l
(%t5)                               A
                                     c i k
\end{lstlisting}

The {\tt itensor} package can also be used to declare symmetry properties of indexed objects and use them to simplify indexed expressions. In the following example, we declare the tensor $\vec{A}$ to be antisymmetric, and then apply this rule to simplify an expression:

\begin{lstlisting}[frame=single,xleftmargin=0.25in,xrightmargin=0.25in]
(%i6) decsym(A,2,0,[anti(all)],[])$
(%i7) ishow(A([a,b],[])+A([b,a],[]))$
(%t7)                             A    + A
                                   b a    a b
(%i8) ishow(canform(%))$
(%t8)                                  0
\end{lstlisting}

Last but not least, the {\tt itensor} package understands covariant differentiation and Christoffel-symbols. For instance,

\begin{lstlisting}[frame=single,xleftmargin=0.25in,xrightmargin=0.25in]
(%i9) ishow(covdiff(A([i,j],[]),k))$
                               %1               %1
(%t9)             - A     ichr2    - A     ichr2    + A
                     i %1      j k    %1 j      i k    i j,k
\end{lstlisting}

These are just some of the core capabilities of the {\tt itensor} package that we can use when we explore a field theory with Maxima.

\section{Maxwell's theory in Maxima}
\label{sec:derive}

Our goal is to carry out the derivation outlined above in Section~\ref{sec:fld} using Maxima, automating the steps as much as possible, while using methods that are generic, and may also be used to study other field theories.

The first task is to load the {\tt itensor} package and configure a metric:
\begin{lstlisting}[frame=single,xleftmargin=0.25in,xrightmargin=0.25in]
  (%i1) load(itensor)$
  (%i2) imetric(g)$
  (%i3) igeowedge_flag:true$
\end{lstlisting}
We also set an internal {\tt itensor} flag, {\tt igeowedge\_flag}, which controls how exterior products and exterior derivatives are computed. As explained in the documentation, this flag is a necessary evil as there are two distinct conventions in the literature: $A_i\wedge A_j=\tfrac{1}{2}(A_iA_j-A_jA_i)$ and $A_i\wedge A_j=A_iA_j-A_jA_i$ are both used in different contexts. The default behavior if {\tt itensor}, with {\tt igeowedge\_flag=false}, is the first of these two cases; we need the second form.

We next define the electromagnetic field tensor $F_{\mu\nu}$ in terms of its components. We can show right away that its exterior derivative is zero, satisfying Gauss's law for magnetism and Faraday's law ``out of the box'', confirming (\ref{eq:Maxwell1}):
\begin{lstlisting}[frame=single,xleftmargin=0.25in,xrightmargin=0.25in]
(%i4) components(F([m,n],[]),extdiff(A([m],[]),n))$
(%i5) extdiff(F([m,n],[]),k);
(%o5)                                  0
\end{lstlisting}
As the name implies, the {\tt extdiff} function calculates the exterior derivative of its argument with respect to an index.

\vbox{
We now write down ${\cal L}_{\rm EM}$:

\begin{lstlisting}[frame=single,xleftmargin=0.25in,xrightmargin=0.25in]
(%i6) L:ishow(-1/4*F([k,l])*F([a,b],[])*g([],[k,a])*g([],[l,b])
              +j([k],[])*A([l],[])*g([],[k,l]))$
                            k a  l b
                           g    g    (A    - A   ) (A    - A   )
               k l                     b,a    a,b    l,k    k,l
(%t6)         g    j  A  - -------------------------------------
                    k  l                     4
\end{lstlisting}}

We can now discard the component definition of $F_{\mu\nu}$ as it has served its purpose, and we would like the final result to appear in terms of $F_{\mu\nu}$ for simplicity. This is not uncommon in complex tensor calculus derivations: we may introduce definitions or simplification rules to serve us during a part of a calculation, but then discard those rules when they are no longer needed.

Instead, we now define the symmetry properties of $F_{\mu\nu}$. We also introduce new simplification rules, one ({\tt Maxwell}) to recognize $F_{\mu\nu}$ when expressed through $A_\mu$, the other to let Maxima know that $F^{\mu\nu}_{;\mu\nu}=0$ on account of the fact that $F_{\mu\nu}$ is antisymmetric:
\begin{lstlisting}[frame=single,xleftmargin=0.25in,xrightmargin=0.25in]
(%i7) remcomps(F)$
(%i8) decsym(F,0,2,[],[anti(all)])$
(%i9) matchdeclare(a,atom,b,atom)$
(%i10) apply(defrule,[Maxwell,extdiff(A([a],[]),b),F([a,b],[])])$
(%i11) defrule(CC,'covdiff('covdiff(F([],[a,b]),b),a),0)$
\end{lstlisting}
Next, we compute one of the terms of the Euler-Lagrange equation, and then lend a hand to Maxima to re-express this result in terms of $F_{\mu\nu}$:
~\par\noindent
\begin{lstlisting}[frame=single,xleftmargin=0.25in,xrightmargin=0.25in]
(%i12) ishow(diff(L,A([m],[],n)))$
          k a  l b                      n       m         m       n
         g    g    (A    - A   ) (kdelta  kdelta  - kdelta  kdelta )
                     b,a    a,b         k       l         k       l
(%t12) - -----------------------------------------------------------
                                      4
                     k a  l b        n       m         m       n
                    g    g    (kdelta  kdelta  - kdelta  kdelta ) (A    - A   )
                                     a       b         a       b    l,k    k,l
                  - -----------------------------------------------------------
                                                 4
(%i13) ishow(canform(contract(expand(apply1(%,Maxwell)))))$
                                       m n
(%t13)                              - F
\end{lstlisting}
~\par
We can now complete the Euler-Lagrange equation by differentiating and adding the remaining term. At this point, we use
the inert form, {\tt 'covdiff}, of the covariant derivative operator,
as we do not require Maxima to resolve covariant derivatives in terms of Christoffel symbols:

\vbox{
\begin{lstlisting}[frame=single,xleftmargin=0.25in,xrightmargin=0.25in]
(%i14) ishow(contract(diff(L,A([m],[])))+'covdiff(-%,n)=0)$
                                    m n        m
(%t14)                     covdiff(F   , n) + j  = 0
\end{lstlisting}}
This confirms (\ref{eq:Maxwell1}).

Finally, we can also show that the current, $j_\mu$, is conserved:
\begin{lstlisting}[frame=single,xleftmargin=0.25in,xrightmargin=0.25in]
(%i15) ishow(apply1(map(lambda([x],'covdiff(x,m)),lhs(%)),CC) = covdiff(rhs(%),m))$
                                       m
(%t15)                        covdiff(j , m) = 0
\end{lstlisting}
This completes our derivation.

In a mere 15 lines of {\tt itensor} code, we were able to derive the generally covariant field equations of electromagnetism and also demonstrate that the 4-current is conserved. In other words, Maxima, with only minimal hand-holding in the form of simplification rules and some clever simplification choices, was able to re-derive the entirety of Maxwell's theory, using just 15 lines of code.

\vbox{
Indeed, here is the raw Maxima code for the above exercise, in its entirety:
\begin{lstlisting}[frame=single,xleftmargin=0.25in,xrightmargin=0.25in]
load(itensor)$
imetric(g)$
igeowedge_flag:true$
components(F([m,n],[]),extdiff(A([m],[]),n))$
extdiff(F([m,n],[]),k);
L:ishow(-1/4*F([k,l])*F([a,b],[])*g([],[k,a])*g([],[l,b])
        +j([k],[])*A([l],[])*g([],[k,l]))$
remcomps(F)$
decsym(F,0,2,[],[anti(all)])$
matchdeclare(a,atom,b,atom)$
apply(defrule,[Maxwell,extdiff(A([a],[]),b),F([a,b],[])])$
defrule(CC,'covdiff('covdiff(F([],[a,b]),b),a),0)$
ishow(diff(L,A([m],[],n)))$
ishow(canform(contract(expand(apply1(%,Maxwell)))))$
ishow(contract(diff(L,A([m],[])))+'covdiff(-%,n)=0)$
ishow(apply1(map(lambda([x],'covdiff(x,m)),lhs(%)),CC) = covdiff(rhs(%),m))$
\end{lstlisting}}

\section{Discussion}
\label{sec:end}

In this short paper, we saw how the {\tt itensor} package of the Maxima computer algebra system can be used effectively by the field theorist to postulate a generally covariant field theory, derive the corresponding field equations, and investigate its properties.


Similar derivations can be carried out for theories that are more complex, including general relativity itself (the derivation, starting with the Einstein--Hilbert Lagrangian density, is in fact one of the example demonstrations supplied with the {\tt itensor} package), modified gravity theories, or other field theories. The capability of Maxima to effectively manipulate functional derivatives is crucial to these exercises.

The features of {\tt itensor} that we used in this derivation only scratch the surface. We should note that {\tt itensor} has capabilities to deal with non-conventional theories such as theories involving nonsymmetric metrics or torsion.

The indicial tensor manipulation capability of {\tt itensor} is directly connected to the component tensor manipulation functionality of another Maxima package, {\tt ctensor}. A conversion function, {\tt ic\_convert}, can translate indexed {\tt itensor} expressions into program snippets for {\tt ctensor}. Thus, it is possible to use the indicial capabilities of {\tt itensor}, as in the example above, to derive a set of field equations, which can then be solved by postulating a specific coordinate system and metric using {\tt ctensor}.

In the end, it is of course necessary as always to ``guide'' the computer algebra system through the steps. Finding the right simplifications is often a trial-and-error process. Furthermore, one should always treat the computer algebra system with a healthy degree of suspicion: No software is perfect, and subtle bugs in its implementation can lead to erroneous results. Therefore, the computer algebra system's most productive use is to verify results. When used as a tool for discovery, any new results should be checked by other means (or, at the very least, by alternative derivations.)

Nonetheless, we find that {\tt itensor} is immensely helpful when it comes to the study of field theories and investigating their properties. We hope that this paper proves useful as it explores the application of {\tt itensor} to the well-known but still nontrivial and highly educational case of Maxwell's field theory of electromagnetism in generally covariant form.

\section*{Acknowledgments}

Maxima in its present form exists thanks to the efforts of its many volunteer developers.

~\par

VTT acknowledges the generous support of Plamen Vasilev and other Patreon patrons.


\bibliography{refs}

\begin{thebibliography}{7}
\expandafter\ifx\csname natexlab\endcsname\relax\def\natexlab#1{#1}\fi
\expandafter\ifx\csname bibnamefont\endcsname\relax
  \def\bibnamefont#1{#1}\fi
\expandafter\ifx\csname bibfnamefont\endcsname\relax
  \def\bibfnamefont#1{#1}\fi
\expandafter\ifx\csname citenamefont\endcsname\relax
  \def\citenamefont#1{#1}\fi
\expandafter\ifx\csname url\endcsname\relax
  \def\url#1{\texttt{#1}}\fi
\expandafter\ifx\csname urlprefix\endcsname\relax\def\urlprefix{URL }\fi
\providecommand{\bibinfo}[2]{#2}
\providecommand{\eprint}[2][]{\url{#2}}

\bibitem[{\citenamefont{Landau and Lifshitz}(1975)}]{LL1975}
\bibinfo{author}{\bibfnamefont{L.~D.} \bibnamefont{Landau}} \bibnamefont{and}
  \bibinfo{author}{\bibfnamefont{E.~M.} \bibnamefont{Lifshitz}},
  \emph{\bibinfo{title}{{Theoretical Physics}}}, vol. \bibinfo{volume}{II:
  Field Theory} (\bibinfo{publisher}{Nauka, Moscow}, \bibinfo{year}{1975}).

\bibitem[{\citenamefont{Lovelock and Rund}(1989)}]{LR1989}
\bibinfo{author}{\bibfnamefont{D.}~\bibnamefont{Lovelock}} \bibnamefont{and}
  \bibinfo{author}{\bibfnamefont{H.}~\bibnamefont{Rund}},
  \emph{\bibinfo{title}{{Tensors, Differential Forms, and Variational
  Principles}}} (\bibinfo{publisher}{Dover Publications},
  \bibinfo{year}{1989}).

\bibitem[{\citenamefont{Fateman}(1979)}]{FATEMAN1979}
\bibinfo{author}{\bibfnamefont{R.~J.} \bibnamefont{Fateman}}, in
  \emph{\bibinfo{booktitle}{Macsyma Users' Conference}} (\bibinfo{year}{1979}),
  \bibinfo{note}{\\\url{https://people.eecs.berkeley.edu/~fateman/papers/simplifier.txt}},
  \urlprefix\url{https://people.eecs.berkeley.edu/~fateman/papers/simplifier.txt}.

\bibitem[{\citenamefont{Moses}(2012)}]{MOSES2012}
\bibinfo{author}{\bibfnamefont{J.}~\bibnamefont{Moses}},
  \bibinfo{journal}{Journal of Symbolic Computation}
  \textbf{\bibinfo{volume}{47}}, \bibinfo{pages}{123} (\bibinfo{year}{2012}),
  ISSN \bibinfo{issn}{0747-7171},
  \urlprefix\url{https://www.sciencedirect.com/science/article/pii/S0747717110001483}.

\bibitem[{\citenamefont{{Bogen} and {Pavelle}}(1977)}]{Bogen1977}
\bibinfo{author}{\bibfnamefont{R.~A.} \bibnamefont{{Bogen}}} \bibnamefont{and}
  \bibinfo{author}{\bibfnamefont{R.}~\bibnamefont{{Pavelle}}},
  \bibinfo{journal}{Letters in Mathematical Physics}
  \textbf{\bibinfo{volume}{2}}, \bibinfo{pages}{55} (\bibinfo{year}{1977}).

\bibitem[{\citenamefont{Toth}(2005)}]{Toth2005}
\bibinfo{author}{\bibfnamefont{V.~T.} \bibnamefont{Toth}},
  \bibinfo{journal}{ArXiv} \textbf{\bibinfo{volume}{cs/0503073}}
  (\bibinfo{year}{2005}).

\bibitem[{\citenamefont{{MacCallum}}(2018)}]{2018LRR}
\bibinfo{author}{\bibfnamefont{M.~A.~H.} \bibnamefont{{MacCallum}}},
  \bibinfo{journal}{Living Reviews in Relativity}
  \textbf{\bibinfo{volume}{21}}, \bibinfo{eid}{6} (\bibinfo{year}{2018}).

\end{thebibliography}


\end{document}